\documentclass[twocolumn,prc,nofootinbib,showpacs,epsfig]{revtex4}
\usepackage{graphicx}
\usepackage{epsfig}
\usepackage{color}
\usepackage{amsmath}
\usepackage{booktabs}
\usepackage{lineno}

\newcommand{\nuc}[2]{$^{#2}\rm #1$}
\newcommand{\gline}{$\gamma$-line}
\newcommand{\glines}{$\gamma$-lines}
\newcommand{\gray}{$\gamma$-ray}
\newcommand{\grays}{$\gamma$-rays}

\newcommand{\baseT}[2]{\mbox{$#1\,\rm{x}\,10^{#2}$}}
\newcommand{\baseTsolo}[1]{$10^{#1}$}

\newcommand{\tab}{{Tab.~}}
\newcommand{\eq}{{Eq.~}}
\newcommand{\fig}{{Fig.~}}

\def\Journal#1#2#3#4{{#1} {\bf #2}, #3 (#4)}

\def\NIMA{{\em Nucl. Instrum. Methods} A}

\def\NPA{{\em Nucl. Phys.} A} 
\def\NP{\em Nucl. Phys.}

\def\JPG{\em Journal of Physics G}

\def\PR{\em Phys. Rev.} 
\def\NAF{\em Z. Naturforschung A} 

\def\PHM{\em Philos. Mag.} 
\def\PRD{{\em Phys. Rev.} D} 
\def\PRC{{\em Phys. Rev.} C} 
\def\CPC{{\em Chin. Phys.} C}

\def\ZP{\em Z. Phys.}

\def\EPA{{\em Europ. Phys. J.} A}

\def\ARI{\em Appl. Rad. Isot.}

\newcommand{\be}{\begin{equation}}
\newcommand{\ee}{\end{equation}}
\def\bea{\begin{eqnarray}} 
\def\eea{\end{eqnarray}}

\newcommand{\zbb}{2\mbox{$\nu\beta\beta$ - decay} }

\newcommand{\vf}{\mbox{$^{50}$V }}
\newcommand{\tf}{\mbox{$^{50}$Ti }} 
\newcommand{\cf}{\mbox{$^{50}$Cr }}

\newcommand{\iso}[2]{{\ensuremath{{}^{#2}\mathrm{#1}}}}

\begin{document}

\flushbottom
\title{A new investigation of half-lives for the decay modes of \vf}

\author{M. Laubenstein$^{a}$} \email{matthias.laubenstein@lngs.infn.it} 
\author{B. Lehnert$^{b}$} \email{bjoernlehnert@lbl.gov}  
\author{S. S. Nagorny$^{c}$} \email{sn65@queensu.ca} 
\author{S. Nisi$^{a}$} \email{stefano.nisi@lngs.infn.it} 
\author{K. Zuber$^{d,e}$} \email{zuber@physik.tu-dresden.de}

\affiliation{
$^{a}$INFN - Laboratori Nazionali del Gran Sasso, 67100 Assergi (AQ), Italy\\
$^{b}$Nuclear Science Division, Lawrence Berkeley National Laboratory, Berkeley, CA 94720, U.S.A.\\
$^{c}$Queen's University, Physics Department, Kingston, ON, K7L 3N6, Canada \\
$^{d}$Institut f\"ur Kern- und Teilchenphysik, TU Dresden, 01069 Dresden, Germany\\
$^{e}$Magyar Tudományos Akadémia Atomki, 4026 Debrecen, Hungary}

%
%
%
%
%
%
%

\begin{abstract}
A new search for the decay modes of the 4-fold forbidden non-unique decay of \vf 
has been performed at the Gran Sasso Underground Laboratory (LNGS). In total
an exposure of 197~kg$\times$d has been accumulated. The half-life for the electron capture
into the first excited state of $^{50}$Ti has been measured with the highest precision to date as \mbox{\baseT{2.67_{-0.18}^{+0.16}}{17}~yr\ (68\%\ C.I.)} in which systematics uncertainties dominate. The search for the $\beta$-decay into the first
excited state of $^{50}$Cr resulted in a lower limit of \mbox{\baseT{1.9}{19}~yr\ (90\%\ C.I.)}, which is an improvement of almost one order of magnitude compared to existing
results. The sensitivity of the new measurement is now in the region of theoretical
predictions.

\end{abstract}


\maketitle

\section{Introduction}

The search for extremely rare events like dark matter, nucleon decays or neutrino-less double beta decay is a wide spread activity in particle astrophysics and for the search 
of physics beyond the Standard Model of particle physics. These experiments are typically located underground to reduce backgrounds from cosmic rays with further
reduction of background by using materials with low radioactive contaminations. As an
example in this way the neutrino-accompanied double beta decay ($2\nu\beta\beta$-decay) has been 
observed for almost a dozen isotopes whose half-life are in the region of 10$^{18-24}$~yr.\\
Having achieved such a sensitivity, it is an obvious step also to study other very long-living nuclides, which are typically highly forbidden beta-decays and electron captures (EC). It turns out that in extremely highly forbidden decays like $^{48}$Ca and $^{96}$Zr
single $\beta$-decay and \zbb compete with each other. For $^{96}$Zr the $\beta$-decay half-life is calculated at $2.4 \times 10^{20}$ yr \cite{hei07}. The 2$\nu\beta\beta$ half-life has
been measured by NEMO-3 to be $T_{1/2} = 2.35 \pm 0.14 \rm(syst.) \pm 0.16 (stat.) \times 10^{19}$ yr
\cite{arg10}. 
Single $\beta$-decay has been searched for and a lower limit of 
$T_{1/2} > 2.4 \times 10^{19}$~yr
has been given \cite{fin16}. 
A similar case can be made for $^{48}$Ca: 
The 2$\nu\beta\beta$ decay has been measured as $T_{1/2} = 6.4^{+0.7}_{-0.6} \rm(stat.) ^{+1.2}_{-0.9} (syst.)
\times 10^{19}$~yr by NEMO-3 \cite{arn16}. 
A half-life limit of the $\beta$-decay to the corresponding $5^+$ excited state of \iso{Ti}{48}
results in a lower limit of $T_{1/2} > 2.5 \times 10^{20}$~yr \cite{bak02}. 
These results slightly indicate that indeed \zbb is more likely
than $\beta$-decay. 

\begin{figure}
  \centering
  \includegraphics[width=0.5\textwidth]{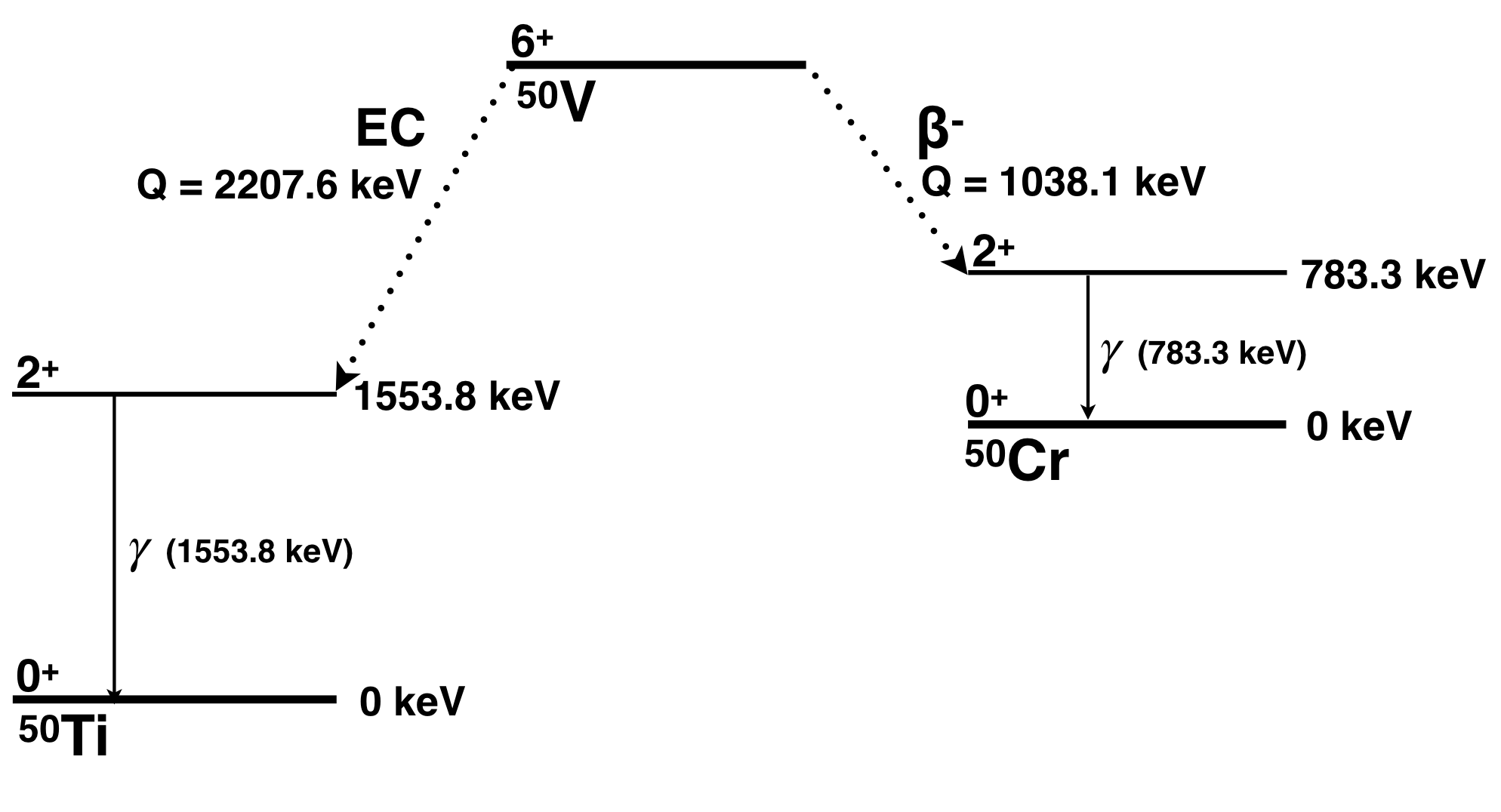}
\caption{Decay scheme of $^{50}$V. Two excited states can be populated, one via EC to \tf under emission of a 1553.77 keV \gray\ and the $\beta$-decay into the first excited state of \cf resulting in a 783.29 keV \gray.}
  \label{pic:levelvf}
\end{figure}

The next group of nuclides, with one unit of spin change less, are 4-fold forbidden non-unique decays ($\Delta I ^{\Delta \pi} = 4^+$) and contain isotopes like
$^{113}$Cd, $^{115}$In and \nuc{V}{50}, the latter being explored in this paper.\\
The isotope \vf is quite unique in the sense that in contrast to $^{113}$Cd and $^{115}$In
the ground state transition is even higher-forbidden, leaving only 4-fold forbidden non-unique decay modes 
into the first excited state of \cf and $^{50}$Ti, both characterized as 6$^+ \rightarrow 2^+$ transitions. 
The ground state transitions to both isotopes are even 6-fold forbidden non-unique decays. The decay
scheme is shown in \fig \ref{pic:levelvf}. 

The Q-value for the $\beta$-decay into \cf is (1038.06~$\pm$~0.30)~keV  and for electron capture (EC) into \tf
it is (2207.6~$\pm$~0.4)~keV, respectively \cite{hua17}. There is only one excited state in each daughter nucleus 
which can be populated. The corresponding \glines\ to search for are 1553.77~keV for EC into the first excited state of \tf and 783.29~keV for the $\beta$-decay into the first excited state of \cf, respectively. The photon emission probability of both E2 transitions is 1 (with a negligible uncertainty). \\ \indent
The history of attempts to observe the decay of \vf is quite special and lasting for more than 60 years where several potential detections were proven wrong by newer, more sensitive experiments. This pattern repeated several times \cite{hei55,glo57,bau58,mcn61,wat62,son66,pap77,alb84,sim85,sim89}. Furthermore, the deduced half-life is different in different articles, while uncertainties claimed were typically well beyond 20\%. 
The measurements within the last 45 years are compiled in \tab \ref{tab:exppara}.
A first clear observation of the EC-branch decay to \tf has been reported by \cite{dom11}. In the meantime a first theoretical nuclear shell model calculation has been performed for the $\beta^-$-decay mode 
which predicts a half-life value of about $2\times 10^{19}$~yr \cite{haa14}.   

The aim of this paper is to perform a high statistics measurement of the  EC-branch decay to \tf and improve and explore the  $\beta$-decay branch of the \vf decay taking advantage of the most sophisticated low background detector system based on HPGe-detectors in one of the deepest underground laboratories available.

\begin{table}[htp]
\begin{center}
\begin{tabular}{ccccc}
\hline
Mass   & Time       & $T_{1/2}^{EC}$       & $T_{1/2}^{\beta^-}$              & Ref., year \\
  \mbox{[g]}   & [d]          & [$10^{17}$~yr]           & [$10^{17}$~yr]                       &     \\
\hline 
4000   &   48.88    &  $>8.8$              & $>7.0$                           & \cite{pap77}, 1977\\ 
4250   &   135.5    &  $1.5^{+0.3}_{-0.7}$ & $>4.3$                           & \cite{alb84}, 1984\\ 
100.6  &   8.054    &  $1.2^{+0.8}_{-0.4}$ & $>1.2$                           & \cite{sim85}, 1985\\ 
337.5  &   46.21    &  $2.05 \pm 0.49$     & $8.2^{+13.1}_{-3.1}$             & \cite{sim89}, 1989\\
255.8  &   97.8     &  $2.29 \pm 0.25$     & $>15$                            & \cite{dom11}, 2011  \\ 
\hline
\end{tabular}
\caption{\label{tab:exppara} Measurements of \vf decays in the last 45 years.}
\end{center}
\end{table}

\section{Experimental Setup}

The vanadium sample was produced from vanadium flakes by multifold electron beam melting (EBM) under high vacuum as described in detail in \cite{azh01}. Vanadium with an initial purity grade of 97 wt\% was used as starting material. The small flakes of vanadium metal were compressed into tablets with a dimension of 30$\times$10 mm (d$\times$h) and about 35 g of mass each. The EBM purification process produced ingots with a diameter of about 45 mm, which were later cut into discs of about $40\times 10$~mm (d$\times$h). 

\subsection{Chemical purity of the vanadium sample}

In order to determine the residual impurities and their concentration and to evaluate the efficiency of the refining process, several analyses have been performed.
A general comparative analysis of elemental impurities in the vanadium before and after purification was carried out by laser mass-spectrometry. Two samples of vanadium in form of small plates $5\times5\times3$~mm$^3$, with chemically cleaned surfaces were analyzed.

The results are shown in \tab \ref{tab:impurities}. As can be seen, the EBM refining method is rather effective for elements that have a high separation factor at the typical temperature for the vanadium refining process of 2400 K. For example, the Cr concentration was reduced by two orders of magnitude, whereas K was reduced 75 times. On the other hand, for some elements (e.g.\ Ni, Si) almost no purification occurred. More details can be found in reference \cite{bob14}, where the separation coefficients  $\alpha_i$ for the main impurity elements in vanadium were determined and investigated.

\begin{table}[htbp]
\begin{center}
\begin{tabular}{lclcl}
\hline
Element  & Before  & After \\
& EBM & EBM\\
\hline
Cr &  850 & 7 \\
K &  130 & 1.7 \\
Al &  120 & 11 \\
Cu &  100 & 18 \\
Fe &  70 & 10 \\
P &  70 & 0.5 \\
Cl & 42 & 4 \\
Si &  17 & 17 \\
Ca & 13 & 5 \\
Ni &  8 & 5 \\
Mg & 4 & 0.7 \\
Ti &  3.5 & 1 \\
Na & 3 & 0.4 \\
Zn & 1.6 & 1\\
Mn &  1.5 & 0.11 \\
\hline
\end{tabular}
\medskip
\caption{\label{tab:impurities} 
Concentration of impurities in the vanadium sample before and after purification with electron beam melting (EBM). Values are given in units of 10$^{-6}$~g/g with estimated uncertainty on the level of 30 \%.}
\end{center}
\end{table}

The gas-forming impurities were also removed with high efficiency by the EBM process. The outgassing of the vanadium samples was measured before, and after purification using an MX7203 mass spectrometer \cite{azh06}, within a temperature range from 25$^{\circ}$C to 800$^{\circ}$ C. Four main components are released from the vanadium samples during the thermal desorption, H$_2$O, CO, N$_2$ and CO$_2$. The intensity of outgassing for the sample after refining is five times less than for the starting metal. Overall, the concentration of oxygen was reduced from $(700 \pm 100) \times 10^{-6}$~g/g in the initial vanadium flakes to 
$(170\pm20)\times 10^{-6}$~g/g, which is the dominant impurity in the refined metal sample. The outgassing impurities as well as the total impurity of the vanadium sample are the lowest achieved so far. The vanadium purity is well determined with ($99.97\pm0.01$)~wt\% which allows to reduce the mass uncertainty in the analysis.

The concentrations of \nuc{Th}{232} and \nuc{U}{238} were measured by High Resolution Inductively Coupled Plasma Mass Spectrometry (HR-ICP-MS), with a Thermo Fisher Scientific ELEMENT2 instrument. The sample was dissolved in acid solution and diluted for the measurement. A semi-quantitative analysis was performed, i.e.\ the instrument was calibrated based on a single reference standard solution of \nuc{Th}{232} and \nuc{U}{238}. The results are shown in \tab \ref{tab:RadImpurities}. The signal for \nuc{U}{238} was close to background, which means that the calculated concentration for this isotope (in the first column of \tab \ref{tab:RadImpurities}) is affected by a large relative uncertainty (30\%).

In order to have a more sensitive measurement of \nuc{Th}{232} and \nuc{U}{238}, another analysis of vanadium was performed, extracting and pre-concentrating the analytes. Chromatographic extraction columns (EXC) packed with U/TEVA\textsuperscript{\textregistered} resins (Triskem International, France) were used for the selective extraction of thorium and uranium from vanadium after dissolution. As shown in \tab \ref{tab:RadImpurities}, the obtained results agree with those obtained from the first analysis. The reliability of the extraction procedure was confirmed by a recovery test: 80\% recovery was found for both elements.

\begin{table}[htbp]
\begin{center}
\begin{tabular}{lclcl}
\hline
Isotope  & Without EXC  &  With EXC \\
\hline
\nuc{Th}{232} &  $<$ 0.5  & $<$ 0.025 \\
\nuc{U}{238} &  0.5 & 0.35\\
\hline
\end{tabular}
\medskip
\caption{\label{tab:RadImpurities}Concentration of \nuc{Th}{232} and \nuc{U}{238} in the vanadium sample before purification obtained by HR-ICP-MS measurements with and without chromatographic extraction (EXC) of the analyses. The concentrations are given in units of 10$^{-9}$~g/g. The estimated relative uncertainties are 30\%.}
\end{center}
\end{table}

\begin{table}[htbp]
\begin{center}
\begin{tabular}{lclcl}
\hline
Isotope  & Abundance in \% &  Abundance in \% \\
&  from \cite{ber11}   & from our  samples \\
\hline
\vf & 0.250 $\pm$ 0.004 & 0.239 $\pm$ 0.012 \\
$^{51}$V & 99.750 $\pm$ 0.004 & 99.761 $\pm$ 0.050\\
\hline
\end{tabular}
\medskip
\caption{\label{tab:isoAbundance}The isotopic composition of the vanadium sample in comparison with literature values 
\cite{ber11}.}
\end{center}
\end{table}

Additional ICP-MS measurements were carried out to confirm the isotopic composition of the vanadium sample which could differ from the literature value e.g.\ through extraction from different geological deposits. The measured isotopic abundances are shown in \tab \ref{tab:isoAbundance} and are consistent with the literature values. Hence, we have no evidence that the vanadium sample has an altered isotopic abundance and we proceed using the more precise literature value in the analysis.

\subsection{Radiopurity of the vanadium sample}

\begin{figure}
\centering
\includegraphics[width=0.4\textwidth]{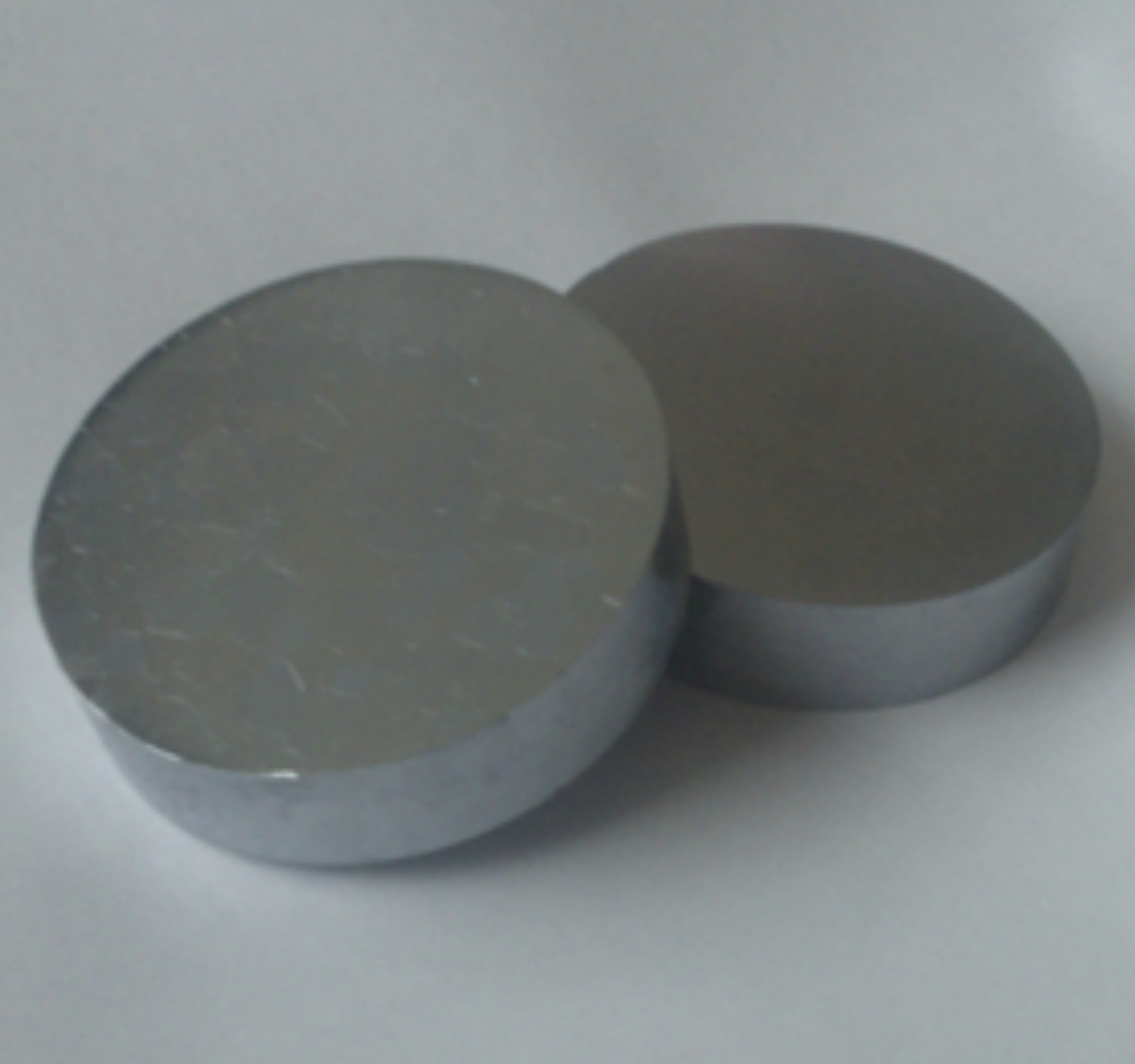}
\caption{%
Two vanadium discs after multifold electron beam melting (EBM).
\label{fig:pic_rawVanadium}}
\end{figure}

\begin{figure}
\centering
\includegraphics[width=0.4\textwidth]{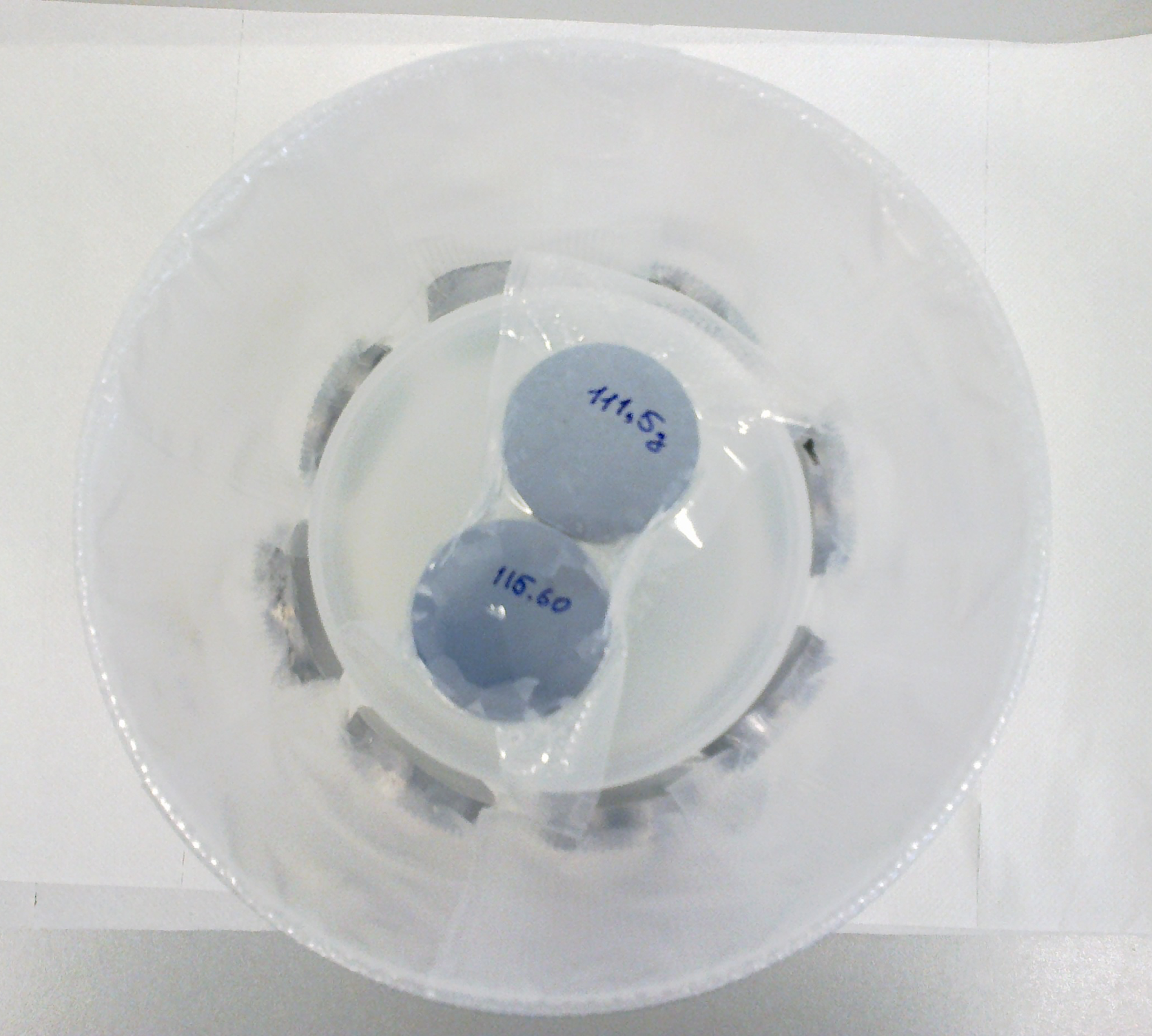}
\caption{%
Arrangement of the ten purified vanadium discs inside a Marinelli beaker ready to be measured.
\label{fig:pic_detAssembly}}
\end{figure}

The initial and the purified metal sample were measured by means of \gray\ spectrometry with ultra-low background high purity germanium (ULB-HPGe) detectors. The measurements were done in the STELLA (SubTerranean Low Level Assay) facility deep underground in the Gran Sasso National Laboratories of the INFN (Italy), details can be found in [17-21].
The sample of initial vanadium with a mass of 987.9 g in form of small metallic flakes was placed in a polypropylene container in Marinelli geometry (GA-MA Associates, type 141G), above the end cap of the ULB-HPGe detector. This initial sample was measured for 35.3~d. 
After the purification by EBM, the vanadium was in the form of cylindrical ingots as shown in \fig \ref{fig:pic_rawVanadium}. The ingots were cut into 10 disks and machined on the outside in order to obtain discs with equal diameter of 40 mm. A total sample mass of 818.5 g was obtained. The surface of the machined vanadium disks was purified by etching with 0.1M HNO$_3$. Each single disk was sealed in a plastic bag. The measurement geometry of the final sample is shown in \fig \ref{fig:pic_detAssembly} and was optimize to yield the best detection efficiency and lowest self-absorption. 
The two largest disks (111.50 g and 115.60 g) are lying on top of the inner part of a Marinelli beaker (type 141G), and the others are hung 87.5 mm from top of the outer wall on its inside around the endcap (75.95 g, 75.15 g, 75.14 g, 66.55 g, 73.80 g, 72.31 g, 76.45 g, and 76.02 g).  The detection efficiencies for the full energy peaks in the sample-detector arrangement were obtained using the Monte-Carlo simulation code MaGe \cite{bos11}, based on the GEANT4 software package.

The vanadium discs were measured after purification with an ULB-HPGe detector for 240.6~d. The spectrum is shown in \fig \ref{fig:pdf_ROI_beta} together with a 70.3~d background spectrum. The 1553.77~keV peak from the \nuc{V}{50} EC decay mode is clearly visible and the most prominent feature in the source spectrum. Compton features of this \gray\ dominate the background below the peak. \\

\begin{figure}
\centering
\includegraphics[width=0.5\textwidth]{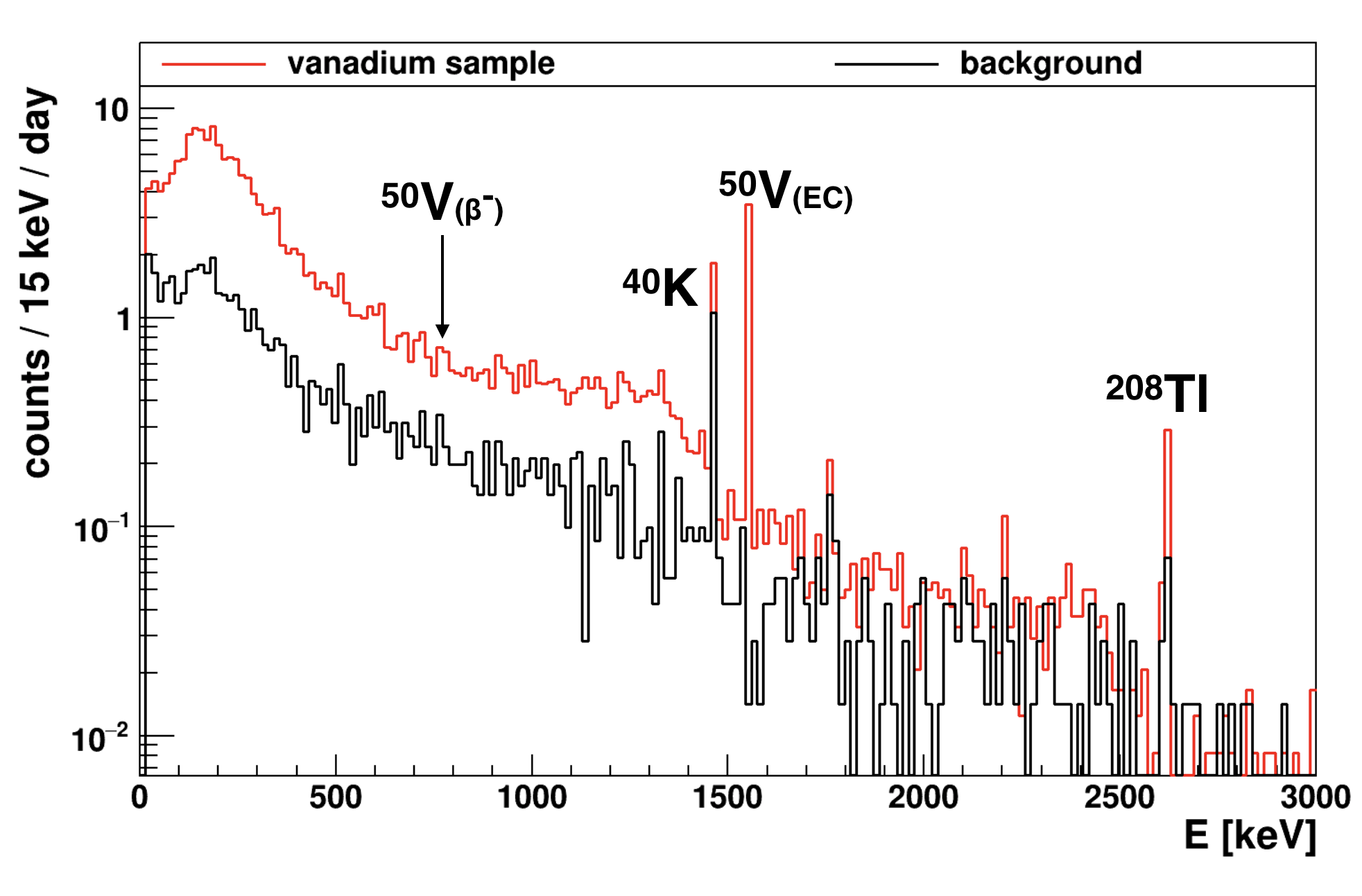}
\caption{%
Spectrum of the vanadium sample (240.6~d) in red and a corresponding background measurement (70.3~d) in black. The two region of interests and prominent background peaks are highlighted.
\label{fig:pdf_ROI_beta}}
\end{figure}

The measured activities of typical background isotopes before and after purification are given in \tab \ref{tab:ScreeningImpurities}. 
In both cases all observed peaks other than the one from \nuc{V}{50} EC are due to \grays\ of the naturally occurring radioactivity coming from the U and Th chains, $^{40}$K, from cosmogenic activation, $^{60}$Co, and from man-made radioactivity, $^{137}$Cs.
As can be seen in \tab \ref{tab:ScreeningImpurities}, there is a significant reduction of the counting rate for all radionuclides after the purification. The activity of $^{40}$K was reduced by a factor of 500. There has been also a significant reduction for the \nuc{U}{238} and \nuc{Th}{232} daughter nuclides, by a factor of 30 and 50, respectively. The background index in the region of interest for the \vf $\beta$-decay (783.29 keV) is reduced from 1.78~cts/(keV$\cdot$kg$\cdot$d) to 0.038~cts/(keV$\cdot$kg$\cdot$d), which enhances the experimental sensitivity by a factor of 7. In vicinity of the 1553.77 keV \gline\ that is emitted in case of the \nuc{V}{50} EC, the background rate has been reduced from 0.172~cts/(keV$\cdot$kg$\cdot$d) to 0.00597~cts/(keV$\cdot$kg$\cdot$d), which improves the sensitivity by a factor of 5. 
A comparison between ICP-MS measurements and \gray\ spectrometry for the vanadium samples before the purification shows that the secular equilibrium in both natural decay chains, uranium and thorium, was broken. This disequilibrium remains also in the \gray\ spectrometry results after the purification.

\begin{table}[ht]
\begin{center}
\begin{tabular}{lll}
\hline
Parent nuclide & A [mBq/kg] & A [mBq/kg] \\
 & before & after \\
\hline
\nuc{Ra}{228} (\nuc{Th}{232})& $41.3\pm1.8$ &$0.6\pm0.1$  \\
\nuc{Th}{228} (\nuc{Th}{232})& $19.8\pm0.8$ &$0.4\pm0.1$  \\
\nuc{Ra}{226} (\nuc{U}{238})& $14.5\pm0.6$ &$0.45\pm0.05$  \\
\nuc{Th}{234} (\nuc{U}{238})& $<92.7$ &$6\pm2$  \\
\nuc{Pa}{234m} (\nuc{U}{238})& $<223$ &$10\pm3$  \\
\nuc{U}{235} & $4.2\pm0.7$ &$0.8\pm0.1$  \\
\nuc{K}{40} & $3460\pm170$ &$7\pm1$  \\
\nuc{Co}{60} & $0.8\pm0.3$ &$<0.1$  \\
\nuc{Cs}{137} & $<0.67$ &$0.06\pm0.02$  \\
\hline
\end{tabular}
\medskip
\caption{\label{tab:ScreeningImpurities} The activity of radionuclides in the vanadium samples before and after purification, obtained by ULB-HPGe measurements. The measurement times are 35.3~d and 240.6~d, respectively. Upper limits are given with 90\% C.L., and the expanded standard uncertainties with k=1.}
\end{center}
\end{table}

\section{Analysis}

The search is separated in two parts investigating the $\beta^-$ and EC decay mode independently. The analysis is based on single \gline\ peak fits at 783.29~keV and 1553.77~keV, respectively, including a semi-empiric background model. The model is composed of a linear function in $\pm20$~keV around the peak of interest and known background \glines\ within this region. Where possible, the strength of these background \glines\ is constrained from more dominant \glines\ elsewhere in the spectrum via prior information in a Bayesian concept. 

The signal counts $s$ in the peak of interest is connected with the half-life $T_{1/2}$ of the decay mode as
\begin{eqnarray}
\label{eq:PdHLtoCounts}
s =
\ln{2} \cdot  \frac{1}{T_{1/2}} \cdot \epsilon \cdot N_A \cdot T \cdot m \cdot f  \cdot \frac{1}{M}\ ,
\end{eqnarray}

where $\epsilon$ is the full energy detection efficiency,
$N_A$ is the Avogadro constant,
$T$ is the live-time (240.6~d), 
$m$ is the mass of the vanadium-sample (818.5~g), 
$f$ is the natural isotopic abundance of \nuc{V}{50} (0.25\%) 
and $M$ the molar mass of natural vanadium (50.94). 
%
The Bayesian Analysis Toolkit (BAT) \cite{Caldwell:2009kh} is used to perform a maximum posterior fit. The likelihood $\mathcal{L}$ is defined as the product of the Poisson probabilities over each bin $i$ for observing $n_{i}$ events while expecting $\lambda_{i}$ events:
\begin{eqnarray}
\mathcal{L}(\mathbf{p}|\mathbf{n}) =
 \prod  \limits_i \frac{\lambda_{i}(\mathbf{p})^{n_{i}}}{n_{i}!} e^{-\lambda_{i}(\mathbf{p})}\ ,
\end{eqnarray}

where \textbf{n} denotes the data and \textbf{p} the set of floating parameters.
$\lambda_{i}$ is taken as the integral of the extended p.d.f.\ $\mathcal{P}$ in this bin
\begin{eqnarray}
\lambda_{i}(\mathbf{p}) &=&
 \int_{\Delta E_{i}} \mathcal{P}(E|\mathbf{p}) dE\ , \label{eq:Ta_expcounts}
\end{eqnarray}

where $\Delta E_{i}$ is the bin width. 
The counts in the fit region used to construct $\mathcal{P}$ are expected from (1) the Gaussian signal peak, (2) the linear background and (3) a set of background peaks:

\begin{eqnarray}
 \label{eq:Ta_pdf}
\mathcal{P}(E|\mathbf{p}) &&=
 \frac{s}{\sqrt{2\pi}\sigma} 
\cdot \exp{\left(-\frac{(E-E_0)^2}{2\sigma^2}\right)}\\[2mm]
&&+  B + C\left( E-E_0 \right) \nonumber\\[2mm]
&&+ \sum_{k } \frac{b_k}{\sqrt{2\pi}\sigma} 
\cdot \exp{\left(-\frac{(E-E_{b_k})^2}{2\sigma^2}\right)}.\nonumber
\end{eqnarray}

The first row is describing the signal peak with the energy resolution $\sigma$ and the \gline\ energy $E_0$ as the mean of the Gaussian.
The second row is describing the linear background with the two parameters $B$\ and $C$.
The third row is describing the $k$ background \glines\ with the strength of the peak $b_{k}$. The specific background \glines\ in the $\beta^-$ and EC mode fit are described further below.

Each free parameter in the fit has a prior associated. The prior for the inverse half-life $(T_{1/2})^{-1}$ is flat. Priors for energy resolution, peak position and detection efficiencies are Gaussian, centred around the mean values of these parameters. The width of these Gaussians are the uncertainty of the parameter values. 
This naturally includes the systematic uncertainty into the fit result.

The uncertainty of the peak positions are set to 0.1~keV.    
The energy scale and resolution is routinely determined using reference point sources including \nuc{Th}{228} and \nuc{Eu}{152}. The main \glines\ of these radionuclides are fitted by a Gaussian distribution and the energy resolution is interpolated by a quadratic function. 
A resolution of $\sigma=0.99$~keV was determined at 1553.77~keV with an estimated uncertainty of 5\%. In the Bayesian framework, the posterior information of the resolution parameter in the fit to the prominent 1553.77~keV \gline\ can be used to update the knowledge of the detector resolution with in-situ data. A posterior resolution of $\sigma=0.92\pm0.02$~keV was determined which is used together with the standard calibration to inform the resolution prior for the $\beta$-decay mode fit. A $\sigma$ of $0.73\pm0.02$~keV is used as prior on the resolution for the peak search at 783.29~keV.

The full energy peak detection efficiencies are determined with Geant4 MC simulations tuned to a calibration standard in the same geometry as the vanadium sample. They are 2.69\% at 783.29~keV and 1.94\% at 1553.77~keV. A 5.0\% uncertainty is assumed based on intercomparison tests and quality checks for single \gray\ emitters. Systematic uncertainties on the measured sample mass (0.01\%), the isotopic abundance (1.6\%), the vanadium concentration in the sample (0.01\%) enter the fit in the same way as the detection efficiency and yield a combined uncertainty on the efficiency parameter as 5.3\%.

The posterior probability distribution is calculated from the likelihood and prior probabilities with BAT. The maximum of the posterior is the best fit. The posterior is marginalized for $(T_{1/2})^{-1}$ which is used to determine the 1$\sigma$ uncertainties defined as the smallest connected 68\% probability region in the distribution. In case the probability distribution significantly includes zero, a lower half-life limit is set with the 90\% quantile of the marginalized $(T_{1/2})^{-1}$ distribution equivalent to the 90\% credibility interval (C.I.).

\subsection{Analysis of $\beta$-decay mode}

The fit is performed in the range between 763.3 and 803.3~keV. Seven known background \glines\ are included in the fit even if they are not clearly visible in the spectrum. The selection is based on an emission probability above 1\% or if the expected count rate is larger than 1 count in the dataset. The background \glines\ are outlined in \tab \ref{tab:BGGammasBetaMode}.

The equilibrium of the \nuc{U}{238} decay chain was found to be broken at \nuc{Ra}{226}. The \gline\ of \nuc{Pa}{234m} at 1001.0 keV (0.84\%) has $37.2 \pm 9.1$ counts which was used together with their respective detection efficiencies to constrain the \nuc{Pa}{234m} \glines\ at 785.96 and 785.4~keV. For the lower \nuc{U}{238} chain, the 609.3~keV \gline\ from \nuc{Bi}{214} (45.5\%) with $45.2 \pm 10.3$ count was used to constrain the 768.36 and 785.96~keV \glines\ from \nuc{Bi}{214} and \nuc{Pb}{214}, respectively.

For the \nuc{Th}{232} chain, the average expectation from the 583.2~keV \glines\ from \nuc{Tl}{208} (30.6\%) with $52.1 \pm 10.0$ counts and from the 911.2~keV \gline\ from \nuc{Ac}{228} (25.8\%) with $51.2 \pm 9.9$ counts was used to constrain the 794.95, 772.29 and 785.96~keV \glines\ in the fit.

\begin{table}[htbp]
\begin{center}
\begin{tabular}{lllll}
\hline
Nuclide & $E$ [keV] & $p_{\rm emit}$ [\%]& prior counts & constraining \gline\ \\
\hline
\nuc{U}{238} chain \\
\nuc{Pa}{234m} & 785.96 & 0.0544 & $2.9\pm0.7$ & \nuc{Pa}{234m} \\
\nuc{Pa}{234m} & 785.4 & 0.317 & $15.9\pm4.0$ & \nuc{Pa}{234m} \\
\nuc{Bi}{214} & 768.36 & 4.89 & $4.3\pm1.0$ & \nuc{Bi}{214} \\
\nuc{Pb}{214} & 785.96 & 1.06 & $1.1\pm0.3$ & \nuc{Bi}{214}\\
\hline
\nuc{Th}{232} chain \\
\nuc{Ac}{228} & 794.95 & 4.25 & $7.5\pm1.5$ & \nuc{Tl}{208} + \nuc{Ac}{228}\\
\nuc{Ac}{228} & 772.29 & 1.49 & $2.5\pm0.5$ & \nuc{Tl}{208} + \nuc{Ac}{228}\\
\nuc{Bi}{212} & 785.96 & 1.102 & $1.9\pm0.4$ & \nuc{Tl}{208} + \nuc{Ac}{228}\\
\hline
\end{tabular}
\medskip
\caption{\label{tab:BGGammasBetaMode} Background \glines\ in the $\beta^-$-decay mode fit window with prior constrained based on more prominent background \glines\ in their respective decay sub-chain. Note that the secular equilibrium of the \nuc{U}{238} decay chain is broken.
}
\end{center}
\end{table}

\begin{figure}
\centering
\includegraphics[width=0.5\textwidth]{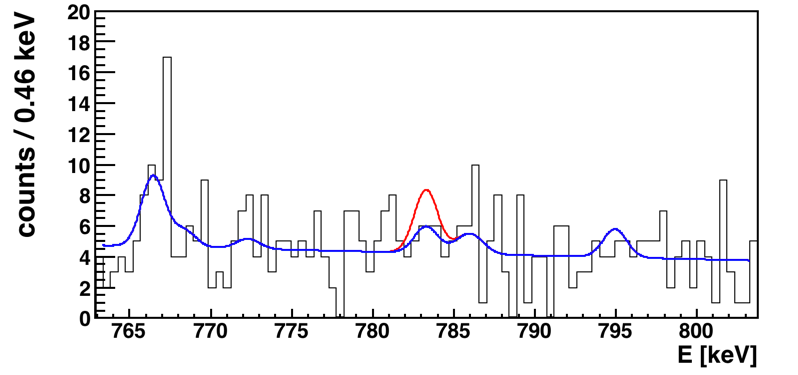}
\caption{%
Spectral fit of the \nuc{V}{50} $\beta^-$ decay \gline\ at 783.29~keV and the background \glines. Shown is the best fit function in blue and the signal process set to the obtained 90\% C.I. limit in red. 
\label{fig:pdf_ROI_beta}}
\end{figure}

The best fit yields a positive signal at $T_{1/2}^{-1} = \baseT{2.27}{-20}$~yr$^{-1}$ or \baseT{4.4}{19}~yr which is distinct from the background only hypothesis or 0~yr$^{-1}$ by 1.2$\sigma$. Hence, no significant signal is observed and the 90\% quantile of the marginalized $T_{1/2}^{-1}$ distribution yields \baseT{5.37}{-20} yr$^{-1}$. This translates into a lower half-life limit of 
$$T_{1/2} (\beta^-) > \rm \baseT{1.9}{19}~yr \ (90\%\ C.I.) \ .$$  

The background below the peak is obtained from fit parameter B in \eq \ref{eq:Ta_pdf} as $0.038\pm0.002$ cts/kg/d.
The fit function is shown in \fig \ref{fig:pdf_ROI_beta} set to the best fit values (blue) and to the 90\% limit of the signal process (red). Systematic uncertainties are included in the result but are negligible compared to the limit of low counting statistics. 
Fixing the peak position, resolution and efficiency priors to their nominal values and repeating the fit without systematic uncertainty changes the limit by $<0.2$\%. Choosing flat priors for the background \glines\ instead of Gaussian constraints changes the limit by 4.1\%.

This result is about one order of magnitude better than the one in \cite{dom11} and approaches the theoretical prediction of $2\times 10^{19}$~yr in \cite{haa14}. The obtained improvement is due to a factor of 7.9 more exposure and a factor of 3.2 lower background level. The analysis method differs with a full spectral fit compared to a counting method. The energy resolution and detection efficiency is similar in both searches. Also note that the limit in \cite{dom11} is based on a 95\% confidence level (C.L.) whereas this result quotes a 90\% credibility interval (C.I.).

\subsection{Analysis of electron capture mode}

The EC \gline\ at 1553.77~keV is clearly visible in the spectrum. 
The fit range is chosen between 1533.8 and 1573.8~keV including two background \glines\ from \nuc{Bi}{214} at 1538.5~keV (0.40\%) and at 1543.3~keV (0.30\%). Constraints for these \glines\ are again taken from the 1764.5~keV \gline\ from \nuc{Bi}{214} ($32.7 \pm 6.6$ counts) resulting in an expectation of $1.4\pm0.3$ and $1.3\pm0.3$ counts, respectively. The expectation varies from the one based on the 609.3~keV \gline\ by about a factor of 5 which is likely due to \nuc{U}{238} background components in different locations which is not completely modeled in the MC. This would result in different attenuation ratios between the two \glines\ in data and MC. Thus, the expectation is taken from the 1764.5~keV \gline\ which is closer in energy to the ROI. 

A \gline\ from \nuc{Pa}{234m} 1553.75  (0.00821\%) overlapping with the signal peak region was constrained with the more prominent \gline\ at 1001.0 keV to 0.3 counts and thus neglected.

The fit finds a best fit value at \baseT{2.67}{17}~yr. The largest connected 68\% interval in the marginalized $T_{1/2}^{-1}$ distribution is \baseT{[2.49-2.83]}{17}~yr
and taken as the uncertainty coming from the fit combining the statistical and systematic uncertainties naturally:
$$T_{1/2} (\rm EC) = \baseT{2.67_{-0.18}^{+0.16}}{17}~yr \ (68\%\ C.I.) \ .  $$

 This uncertainty is 6.4\% of the best fit value and is dominated by the systematic contributions to the efficiency parameter as outlined in the uncertainty budget in \tab \ref{tab:Systematics}. Switching off these systematic uncertainties in the fit, a range of \baseT{[2.58-2.76]}{17}~yr is obtained which is about 3.4\% coming from statistics and energy scale combined. Changing the Gaussian priors for the background peaks to flat priors has no noticeable effect. 

The measured half-life is about 14\% higher and a factor 1.4 more precise compared to \baseT{2.29 \pm 0.25}{17}~yr previously reported in \cite{dom11}. Both measurement agree within uncertainties.

\begin{table}[htbp]
\begin{center}
\begin{tabular}{ll}
\hline
Uncertainty & fraction in $T_{1/2}$  \\
\hline
Isotopic abundance \nuc{V}{50}& 1.6\% \\
Vanadium concentration & 0.1\% \\
Sample mass & 0.01\%  \\
Detection efficiency & 5.0\%  \\
\hline
Subtotal &  5.3\%\\
\hline
Statistics and energy scale & 3.5\% \\
\hline
Fit total   &6.3\% \\
\hline

\end{tabular}
\medskip
\caption{\label{tab:Systematics} Uncertainty budget for the EC half-life measurement. 
}
\end{center}
\end{table}

\begin{figure}[ht]
\centering
\includegraphics[width=0.5\textwidth]{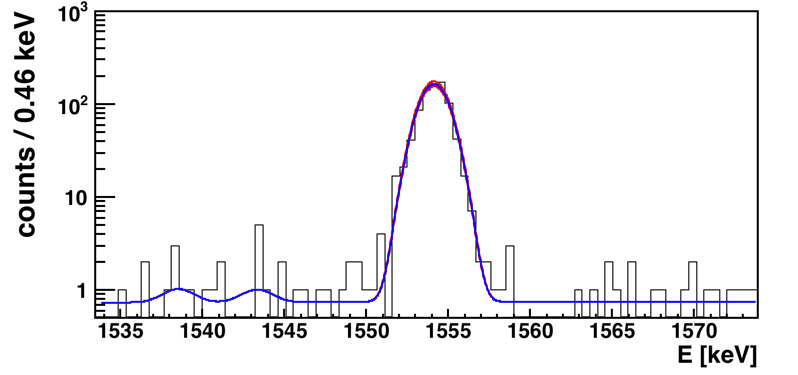}
\caption{%
Spectral fit of the \nuc{V}{50} EC \gline\ at 1553.77~keV and the background \glines. Shown is the best fit function in blue and the signal process set to the $\pm1\sigma$ fit uncertainties in red. 
\label{fig:pdf_ROI_EC}}
\end{figure}

\section{Conclusions}

The 4-fold non-unique forbidden $\beta$-decay and electron capture of \nuc{V}{50} have been investigated with a state of the art ultra low background HPGe setup at LNGS, Italy. The half-life of the EC mode has been determined with unprecedented precision as
\mbox{\baseT{2.67_{-0.18}^{+0.16}}{17}~yr\ (68\%\ C.I.)}. 
The improvement could be achieved with about 10 times higher peak counts compared to a previous measurement which renders the statistical uncertainty subdominant compared to systematic uncertainties. Future improvement can only be expected with a more sophisticated detector calibration or a detector setup with detection efficiency close to 1.

The half-life limit on the $\beta$-decay mode has been improved by more than an order of magnitude to  
\mbox{\baseT{1.9}{19}~yr\ (90\%\ C.I.)}. 
The improvement was mainly possible due to a successful purification of the vanadium sample with electron beam melting in combination with the ultra low background detector deep underground. A longer measurement time and more sample mass compared the the previous best measurement also improved the limit.
The half-life limit of the $\beta$-decay mode is at the point of theoretical predictions at \baseT{2.0}{19}~yr. Even a modest improvement of the half-life sensitivity will either discover the decay or constrain nuclear model calculations.
Further improvements with the setup at hand can be achieved with longer measuring time, more sample mass or enrichment. A significant decrease of the background for the $\beta$-decay search is not easily feasible since the region around 783.3~keV is already dominated by the Compton features of the 1553.8~keV \gline\ coming from the EC mode of the same isotope. A conceptually different detector setup with active Compton detection would be required to reduce the background further. 

Indeed, an approach using vanadium-based (YVO$_4$) crystals as cryogenic scintillating bolometers is discussed in \cite{Pattavina:2018}. First results of the crystal characterization show excellent bolometric performance and light output.
An innovative approach for an efficient detection of the characteristic de-excitation \grays\ following the \nuc{V}{50} $\beta$-decay using triple-coincidences which yields experimental half-life sensitivities at the level of \baseTsolo{20}~yr is proposed as well.
Therefore, the production of high radiopurity YVO$_4$ crystals from EBM purified vanadium which are operated as
scintillating bolometers, read out with auxiliary light detectors and
surrounded by TeO$_2$ bolometers as Compton vetoes are considered as our further
steps.


\end{document}